\documentstyle[12pt]{article}
\def\({\left(}
\def\){\right)}
\def\n{n_{i}}

\newcommand{\be}{\begin{equation}}
\newcommand{\ee}{\end{equation}}
\begin{document}

\centerline{\Large {\bf Statistics of Mass Aggregation in a}} 
\centerline{\Large {\bf Self-Gravitating One-Dimensional Gas} } \vskip 1cm
\centerline{J.C. Bonvin and Ph.A. Martin} \vskip 0.2cm
\centerline{\normalsize {Institut de Physique Th\'eorique, Ecole Polytechnique
 F\'ed\'erale de Lausanne}} 
\centerline{\normalsize{CH - 1015, Lausanne, Switzerland; e-mail: phmartin@sdpdec9.epfl.ch}}
\vskip 1cm
\centerline{J. Piasecki }
\vskip 0.2cm
\centerline{\normalsize{Institute of Theoretical Physics, University of Warsaw}
 }
 \centerline{\normalsize{Ho\.za 69, 00681 Warsaw, Poland}}
\vskip 1cm
\centerline{X. Zotos} 
\centerline{\normalsize{Institut Romand de Recherche
Num\'erique en Physique des Mat\'eriaux (IRRMA)}}
\centerline{\normalsize{EPFL-PPH, Ecublens, CH-1015, Switzerland}} 
\vskip 2cm
\centerline{\bf Abstract}
\vskip 0.5cm

We study at the microscopic level the dynamics of a one-dimensional gravitationally
interacting sticky gas. Initially, $N$ identical particles of mass $m$ with uncorrelated, randomly
distributed velocities fill homogeneously a finite region of space. It is proved that at a
characteristic time a single macroscopic mass is formed with certainty, surrounded by a dust of non
extensive fragments. In the continuum limit this corresponds to a single shock creating a singular
mass density. The statistics of the remaining fragments obeys the Poisson law at all times
following the shock. Numerical simulations indicate that up to the moment of macroscopic aggregation
the system remains internally homogeneous. At the short time scale a rapid decrease in the kinetic
energy is observed, accompanied by the formation of a number $\sim\sqrt{N}$ of aggregates with
masses $\sim m\sqrt{N}$.
\newpage

{\bf 1. INTRODUCTION}\\[0.5cm]

The model of gravitationally interacting sticky particles has been proposed by Zeldovich
[1], and then extensively studied in connection with the problem of large scale structures in
the universe (for a recent review see [2]). The effort has been concentrated on the study of solutions
of the corresponding self-consistent hydrodynamic equations in order to understand how small density
fluctuations arround a homogeneous state could induce observed mass distributions. 

In this paper we analyze the dynamics of a one-dimensional system with an initial homogeneous mass
distribution filling only a finite region of space. As the one-dimensional gravitational
interaction is confining, the whole system will eventually form a single
mass. We start from the microscopic dynamics of  $N$ point particles with randomly chosen initial
velocities. Then we determine the evolution of the statistical distribution of masses formed by
merging at binary sticky collisions. It turns out that a macroscopic mass is formed after a
characteristic finite time with probability one,
surrounded by a cloud of non-extensive fragments. In the continuum limit, this
corresponds to a solution of the system of mass and momentum conservation laws, showing a single shock
at the characteristic time. We emphasize that energy is not conserved in this process.
We thus find not only a simple realization of a particular solution
belonging to the general class studied in [3], but also a detailed description of the statistics
of masses in the course of time.  

As mentioned above, the system is composed of 
point particles attracting each other with  forces
proportional to the product of their masses, and independent of the
interparticle distance.  The corresponding $n$-body Hamiltonian has the form
\begin{equation}
H_{n} = \sum_{i=1}^{n}\frac{p_{i}^{2}}{2m_{i}} + \gamma \sum_{i \;\; <}^{n} 
\sum_{j}^{n}m_{i}m_{j}|x_{i} - x_{j}| , \label{Ham}
\end{equation}
where $\gamma$ is the gravitational constant, 
and $m_{i}, x_{i}, p_{i}$ denote the mass, the position and the momentum of 
particle $i$, respectively.

The Hamiltonian  (\ref{Ham}) with an appropriate value of $n$ defines the 
dynamics of the system in the
time intervals separating  binary collisions. The collisions, which are 
supposed to be perfectly inelastic, are responsible for the mass aggregation. 
When two neighbouring particles $i$ and $j$ meet, they merge instantaneously 
forming a single mass $(m_{i}+m_{j})$ which continues the
motion acquiring at the moment of impact the momentum $(p_{i}+p_{j})$. 
The complete dynamics, involving the aggregation process, is thus subject to 
the mass and momentum conservation laws. On the other hand, the number of 
particles decreases monotonically in the course of time, each collision
replacing two particles by one. 

In a previous work [4] we concentrated on the determination of the
probability $P_{N}(t)$ of merging before time $t$ of the initial $N$-particle 
dust into a single body (we denote here by $N$ the number of particles at time
$t=0$). The rigorous results have been obtained by assuming a uniform 
equidistant configuration of the initial identical masses $m$, characterized
by a constant mass density
\begin{equation}
\rho = \frac{m}{a}, \label{rho}
\end{equation} 
where $a$ is the distance between the nearest neighbours.
The velocities of the particles were supposed to be uncorrelated at $t=0$, 
distributed according to some probability density $\phi (v)$. In particular,
the Gaussian law
\begin{equation}
\phi_{\lambda}(v) = \frac{1}{\sqrt{2\pi}\lambda}{\rm exp}\(- \frac{v^{2}}
{2\lambda^{2}}\) \label{phi}
\end{equation}
has been considered in a detailed way.

It has been found that for a macroscopic amount of matter
$N \rightarrow \infty$ the probability  $P_{N}(t)$ of such a complete
merging vanished for times shorter that the characteristic time
\begin{equation}
t^{*} = \frac{1}{\sqrt{\gamma \rho}} \label{star}
\end{equation}

More precisely, expressing the probability $P_{N}(t)$ in terms of the
relevant time variable
\begin{equation}
\tau = \gamma t - \frac{1}{\rho t} \label{tau}
\end{equation}
we derived a general (independent of the form of distribution $\phi$), 
remarkably simple result
\begin{equation}
P_{N}(\tau = 0) = \frac{1}{N} \label{tauzero}
 \end{equation}
from which the above assertion followed.
 
The exact form of the monotonically increasing function $P_{N}(\tau)$ 
was then determined in the limit of a continuous initial mass distribution.
One way of defining this limit is given below:
\begin{equation}
N \rightarrow \infty , \;\;\; m \rightarrow 0, \;\;\; a \rightarrow 0 
\label{continuum}
\end{equation}
\[ M_{tot} = Nm = {\rm const}, \;\;\; \rho = \frac{m}{a} = {\rm const}, 
\;\;\; \lambda = {\rm const} \]

However, it has been stressed in our concluding remarks that the really
relevant question was that of the macroscopic aggregation, which is quite
different from the complete merging into a single mass. Indeed,
from the physical point of view the important event is certainly that of the 
formation of a macroscopic mass, representing a finite fraction of the total 
mass $M_{tot}=Nm$ in the continuum limit (\ref{continuum}). And this event, 
according to numerical simulations and preliminary analytical results, 
seemed to
occur in the immediate vicinity of the characteristic time $t^{*}$, which 
could not be predicted from the knowledge of the probability $P_{N}(\tau)$.
The main object of the present study is to clarify this point by analyzing
the evolution of the mass distribution in the course of time. In other
words, we determine here the statistics of the mass aggregation providing
a proof that the mass density in the continuum limit (\ref{continuum}) becomes
singular after a finite time $t^{*}$, and takes the form of the Dirac $\delta$
centered on a position of a single macroscopic mass.

The concentration of the mass density on a single point at time $t^{*}$ 
can be simply demonstrated in the special case of a static initial
condition, when all the particles at $t=0$ are at rest: 
$\phi (v)= \delta (v)$. Indeed, suppose that the initial positions of the
particles are $x_{j}(0)=ja, \;\; j=1,2,......,N$. Then, at time $t>0$, the
distances between neighbouring particles shrink to $(a - m\gamma t^{2})$, 
so that all the particles merge simultaneously at the moment $t = t^{*}= 
\sqrt{a/\gamma m}$. The initial mass density uniformly distributed within the
interval $0 < x < Na$ is given by 
\[ \rho (x;0)= \theta (x)\theta (Na-x) \frac{m}{a} =
\theta (x)\theta \(\frac{M_{tot}}{\rho}-x\)\rho , \] 
where $\theta (x)$ is the Heaviside unit step function.
For times $0<t<t^{*}$, it acquires in the continuum limit (\ref{continuum}) 
the value
\begin{equation}
\rho (x;t) = \theta \left[x-\gamma M_{tot}\frac{t^{2}}{2}\right] 
\theta \left[\frac{M_{tot}}{\rho}-\gamma M_{tot}\frac{t^{2}}{2}-x\right]\frac{\rho}
{1 - \rho\gamma t^{2}}  \label{static}
\end{equation}
Equation (\ref{static}) implies the expected result
\begin{equation}
\lim_{t\to t^{*}} \rho (x;t) = M_{tot} \delta \left(x - \frac{M_{tot}}{2\rho}\right)
\label{singular}
\end{equation}
It is shown in the present paper how the above result can be generalized to the
case of a random (Gaussian) initial velocity distribution (\ref{phi}),
reflecting the creation of a macroscopic body at $t=t^{*}$. Moreover, we derive
an analytic formula for the distribution of the remaining nonextensive
microscopic masses.
Computer simulations supplement our analytic approach on two points. They permit to determine the
scaling laws to the continuum limit and to predict the mass distribution before $t^{*}$, a problem
which remains open for the mathematical analysis.

Before closing these introductory remarks let us
recall that the system of mass and momentum conservation laws governing the dynamics of 
one-dimensional aggregation has been recently studied from the point of view of the existence  of
global weak solutions [3]. In their proof the authors have analyzed the dynamics in terms of the
center of mass trajectories  in much the same way as it had already been done in our previously
published  papers [5,4].\\[1cm]

{\bf 2. EVALUATING THE MASS DENSITY}\\[0.7cm]
At the initial moment $t=0$, $N$ identical masses $m$ start the motion with
uncorrelated velocities, distributed according to the probability density
$\phi (v)$. Particle $j$ begins to move from the point $ja,\;\; j=1,2,...,N$.
In order to study the evolution of the mass density one has to determine
the probability density for finding at time $t>0$ a mass $M$ at point
$X$. The dynamics of the system implies that $M$ results from
aggregation of some cluster of neighbouring initial masses, and that $X$
must coincide with the position of the center of mass of the cluster at time 
$t$. So, let us consider a n-particle cluster
\begin{equation}
 (j+1, j+2, ....... , j+n) \label{jn}
\end{equation}

In accordance with the dynamics induced by the Hamiltonian (\ref{Ham}) 
its center of mass $X^{n}_{j+1}$ follows the trajectory
\begin{equation}
X^{n}_{j+1}(t)=\left[ j + \frac{(n+1)}{2}\right] a + \frac{t}{n}\sum_{s=1}^{n}v_{j+s} +
\gamma m(N-n-2j)\frac{t^{2}}{2} \label{cmnp}
\end{equation}
with velocity
\begin{equation}
V^{n}_{j+1}(t)=\frac{1}{n}\sum_{s=1}^{n}v_{j+s} + \gamma m(N-n-2j)t \label{cmnv}
\end{equation}

The n particles (\ref{jn}) merge into a single mass $nm$ before time $t$
if and only if
\begin{equation}
X^{r}_{j+1}(t) > X^{n-r}_{j+r+1}(t),\;\;\; r=1,2,...,n-1 \label{merge}
\end{equation} 
The inequalities (\ref{merge}) express the requirement that for any partition
of the n-particle cluster (\ref{jn}) into subclusters $(j+1,j+2,...,j+r)$ and
$(j+r+1, j+r+2,...,n)$ the centers of mass of the subclusters cross before
time $t$ leading to the total merging. 
In order to guarantee that a single mass $M_{n}=nm$ is actually
observed at the point (\ref{cmnp}) we have still to rule out the possibility
of disturbance which would cause collisions with surrounding masses, 
built up from the initial clusters
\begin{equation}
(1,2,....,j) \label{1j}
\end{equation}
and
\begin{equation}
(n+1,n+2,....,N) \label{nN}
\end{equation}
The unperturbed aggregation of the $n$-particle cluster (\ref{jn}) occurs if
and only if
\begin{equation}
X^{s}_{j-s+1}(t) < X^{n}_{j+1}(t),\;\;\; s=1,2,...,j \label{1jc}
\end{equation}
\begin{equation}
X^{n}_{j+1}(t) < X^{s}_{n+1}(t),\;\;\; s=1,2,...,N-j-n \label{nNc}
\end{equation}
We continue to use here the notation $X^{n}_{i+1}$ to denote the position of
the center of mass  of the n-particle cluster composed of particles 
$(i+1,i+2....,i+n)$.

The inequalities (\ref{1jc}) express the fact that the center of mass
trajectories of the $s$-particle clusters \[(j-s+1,j-s+2,...,j),\;\;\;
s=1,2,...,j\] stay to the left of 
the trajectory $X^{n}_{j+1}(t)$ up to time $t$, and thus do not cross it.
Similarly, the inequalities (\ref{nNc}) exclude crossing of the trajectory 
of the aggregating mass $M_{n}$ with the center of mass trajectories 
on which evolve masses (\ref{nN}) initially to the right of it. 
The necessary and sufficient character of the conditions (\ref{1jc}) and 
(\ref{nNc}) follows from the remark that the dynamics excludes more than 
one crossing between the particle trajectories. 

The probability density for finding at time $t>0$ a mass $M$ at point
$X$ with velocity $V$ can now be written in the form

$$
< \sum_{n=1}^{N}\sum_{j=0}^{N-n}\delta [X-X^{n}_{j+1}(t)]
\delta [V-V^{n}_{j+1}(t)] \delta (M-nm)\prod_{r=1}^{n-1}\theta 
[X^{r}_{j+1}(t)-X^{n-r}_{j+r+1}(t)] 
$$
\be \times\prod_{s=1}^{j}\theta [X^{n}_{j+1}(t)-X^{s}_{j-s+1}(t)] 
\prod^{{\rm N}-j-n}_{s=1}\theta [ X^{s}_{n+1}(t)-X^{n}_{j+1}(t)] >\label{formul} \ee 
where $<...>$ denotes the mean value with respect to the initial velocity
distribution
\begin{equation}
\prod_{i=1}^{{\rm N}}\phi (v_{i}) \label{phiN}
\end{equation}
Multiplying formula (\ref{formul}) by $M$, and integrating over all possible 
masses and velocities we arrive at the expression for the mass density
$$
\rho (X;t) = \sum_{n=1}^{\rm N}\sum_{j=0}^{{\rm N}-n}nm <\delta 
[X-X^{n}_{j+1}(t)]
\prod_{r=1}^{n-1}\theta [X^{r}_{j+1}(t)-X^{n-r}_{j+r+1}(t)]> \label{rhot}
$$
\be\times <\prod_{s=1}^{j}\theta [X-X^{s}_{j-s+1}(t)]> 
<\prod^{{\rm N}-j-n}_{s=1}\theta [ X^{s}_{n+1}(t)-X]>\label{rhot} \ee
In writing equation (\ref{rhot}) we took into account the absence of
correlations in the velocity distribution (\ref{phiN}). As a result
the mean value appearing in equation (\ref{formul}) factorized out into
the product of three averages corresponding to disjoint groups of
velocity variables $(v_{1},...,v_{j}),\;\;(v_{j+1},...,v_{j+n})$
and  $(v_{j+n+1},...,v_{{\rm N}})$.
The calculations greatly simplify in the case of the Gaussian form (\ref{phi})
which will be used in the sequel. In order to show the kind of problems one
has to deal with let us consider in equation (\ref{rhot}) the integration over 
the velocities $(v_{j+1}, v_{j+2},...,v_{j+n})$ of the aggregating n-particle 
cluster. Using  equation (\ref{cmnp}) we find
$$
<\delta [X-X^{n}_{j+1}(t)]
\prod_{r=1}^{n-1}\theta [X^{r}_{j+1}(t)-X^{n-r}_{j+r+1}(t)]>   
$$
$$ =\;\;\int dv_{j+1}...\int dv_{j+n} \phi_{\lambda}(v_{j+1})...\phi_{\lambda}
(v_{j+n})$$
$$ \delta \left[ X-(j+\frac{(n+1)}{2} )a-\frac{t}{n}\sum_{s=1}^{n}v_{j+s}-
\gamma m(N-n-2j)\frac{t^{2}}{2}\right] $$
\be\times \prod_{r=1}^{n-1}\theta \left\{ \sum_{s=1}^{r}v_{j+s}-\frac{r}{n}
\sum_{s=1}^{n}v_{j+s}+\frac{r(n-r)}{2}m\tau \right\}\label{int} \ee 

By introducing new integration variables
\begin{eqnarray}
u_{r} &=& \frac{1}{\lambda}\left( \sum_{s=1}^{r}v_{j+s}-\frac{r}{n}
\sum_{s=1}^{n}v_{j+s}\right) ,\;\; r=1,2,...,n-1\nonumber\\
u_{n}& =& \frac{1}{\lambda}\sum_{s=1}^{n}v_{j+s} 
\label{change}
\end{eqnarray}
one eventually rewrites formula (\ref{int}) in the form
\begin{equation}
\frac{1}{\lambda t}\sqrt{\frac{n}{2\pi}}
{\rm exp}\left[-\frac{n}{2\lambda^{2}}\left( \frac{1}{t} \left[X - 
\frac{M_{tot}+m}{2\rho}\right]-m\tau \left[\frac{N-n}{2}-j\right] \right)^{2} \right] 
P_{n}(\tau)
\end{equation}
Here
$$
P_{n}(\tau) = \sqrt{2\pi n}\int du_{1}...\int du_{n-1}
\phi (u_{1})\phi (u_{2}-u_{1})...\phi (u_{n-1}-u_{n-2})\phi (-u_{n-1})
$$
\be\times  \prod_{r=1}^{n-1}\theta\left[ u_{r}+\frac{r(n-r)m\tau}{2\lambda} \right]
\label{P}\ee
is the probability of merging of an isolated
$n$-particle cluster into a single mass before time $t$, with
\be
\phi (u)=\frac{1}{2\pi}\exp\left( -\frac{u^{2}}{2}\right)
\label{max}
\ee
It is exactly this
quantity which was the main object of our previous study [4]. In particular
the important relation (\ref{tauzero}) has been derived therein.

In quite a similar way one can analyze the other two averages appearing in 
equation (\ref{rhot}). We give the results here only for $t=t^{*}$ (or
$\tau =0$), as our aim is to find the mass density at this characteristic
moment. A straightforward calculation yields then the formulae
\begin{equation}
<\prod_{s=1}^{j}\theta [X-X^{s}_{j-s+1}(t^{*})]>\; =\; B_{j} \left( 
\frac{1}{\lambda t^{*}}\left[X-\frac{M_{tot}+m}{2\rho}\right] \right) 
 \label{chil}
\end{equation}
\begin{equation}
<\prod^{{\rm N}-j-n}_{s=1}\theta [ X^{s}_{n+1}(t^{*})-X]>\; =\;
B_{N-n-j} \left(- \frac{1}{\lambda t^{*}}\left[X-\frac{M_{tot}+m}{2\rho}\right] \right) 
\label{chir}
\end{equation}
where
$$
B_{j} (u) = \int du_{1}...\int du_{j} \phi (u_{1})\phi (u_{2}-u_{1})...
\phi (u_{j}-u_{j-1} )\prod_{s=1}^{j}\theta (su + u_{s}) 
$$
\be \equiv E_{W}[u_{s}> -su,\; s=1,...,j] \label{chij}\ee
As in [4], $u_{s}$ is interpreted as a Brownian path at discrete times $s=1,2,\ldots,$ and the
traditional notation $E_{W}[...]$ for the Wiener measure has also been used here.
Taking into account the relation (\ref{tauzero}) we eventually find that the
mass density at the gravitationally imposed finite time scale $t^{*}$ in the
continuum limit (\ref{continuum}) equals
$$
\rho (X;t^{*})= M_{tot} \lim_{N\to \infty}
\frac{1}{N}\sum_{n=1}^{N}\sum_{j=0}^{N-n}
\frac{1}{\lambda t^{*}}\sqrt{\frac{n}{2\pi}}{\rm exp}
\left[ -\frac{n}{2(\lambda t^{*})^{2}}\(X-\frac{M_{tot}}{2\rho}\)^{2} \right]
\label{rho*}
$$
\be\times B_{j} \left( 
\frac{1}{\lambda t^{*}}\left[X-\frac{M_{tot}}{2\rho}\right] \right)
B_{N-n-j} \left(- 
\frac{1}{\lambda t^{*}}\left[X-\frac{M_{tot}}{2\rho}\right] \right)\label{rho*} \ee

For symmetry reasons one can expect the macroscopic aggregation to take place
at the central point  of the originally uniform 
system. And indeed, it turns out that the mass density (\ref{rho*}) becomes 
singular at this point in the continuum limit (\ref{continuum}). In order to
prove it we shall use now
the Sparre Andersen theorem (Section XII.7 in [6]) which permits to 
determine the generating function\footnote{$B_{n}(u)$ here is the same as $B_{n} $ in (42) of 
[4] with the variable $u$ playing the role of $\tau$. Note that there is a minus sign missing in the
equalities (46) and (48) of [4]. }  
\begin{equation}
p(z;u) = 1 + \sum_{n=1}^{\infty}z^{n}B_{n}(u) 
\end{equation}
One finds (see the discussion in section 5 in [4])
\begin{equation}
p(z;u) = {\exp }\left(\sum_{n=1}^{\infty}\frac{z^{n}}{n}E_{W}[u_{n}>-nu]\right)
\end{equation}
It follows that
$$
p(z;u)p(z;-u)= {\exp }\left(\sum_{n=1}^{\infty}\frac{z^{n}}{n}(E_{W}[u_{n}>-nu]
+ E_{W}[u_{n}>nu]\right) \label{SA}
$$
\be = {\rm exp }\left( \sum_{n=1}^{\infty}\frac{z^{n}}{n}\right) = 
\frac{1}{1-z} \label{SA}\ee
Moreover, it follows from (\ref{SA}) that (setting $B_{0}(u)=1$)
$$
\sum_{n=0}^{\infty}z^{n}B_{n}(u)\sum_{\ell=0}^{\infty}z^{\ell}B_{\ell}(u)=
\sum_{N=0}^{\infty}z^{N}\sum_{\ell=0}^{N}B_{\ell}(u)B_{N-\ell}(-u) =\frac{1}{1-z}  
$$which implies
\begin{equation}
\sum_{j=0}^{N-n}B_{j}(u)B_{N-n-j}(-u) = 1 \label{id}
\end{equation}
Thus the equation (\ref{rho*}) simplifies in a remarkable way to the
form
\begin{equation}
\rho (X;t^{*})= M_{tot}\lim_{N\to \infty}
\frac{1}{N}\sum_{n=1}^{N}\frac{1}{\lambda t^{*}}
\sqrt{\frac{n}{2\pi}}{\rm exp}\left[ -\frac{n}{2(\lambda t^{*})^{2}}
\left(X-\frac{M_{tot}}{2\rho}\right)^{2} \right] \label{rho*00}
\end{equation}
\vspace{3mm}
The main conclusion from equation (\ref{rho*00}) is that at the moment 
$t^{*}$  the mass density 
concentrates at the central point of the system  $X=M_{tot}/2\rho $.
Indeed, if f(X) is a continuous fonction,
$$
\int  f(X)\rho (X;t^{*})dX
$$
\begin{eqnarray}
&=&M_{tot}\lim_{N\to \infty}
\frac{1}{N}\sum_{n=1}^{N}\int dY f\left(\frac{M_{tot}}{2\rho}+\frac{\lambda t^{*}}{\sqrt{n}}Y\right)
\frac{1}{\sqrt{2\pi}}\exp  \left(-\frac{Y^{2}}{2}\right)\nonumber\\
&=&M_{tot}f\left(\frac{M_{tot}}{2\rho}\right)\nonumber
\end{eqnarray}
leading to
\begin{equation}
\rho (X;t^{*}) = M_{tot} \delta \left(X - \frac{M_{tot}}{2\rho} \right) \label{result}
\end{equation}
In particular, the formula
\begin{equation}
\rho (\frac{M_{tot}}{2\rho};t^{*})\sim M_{tot}
\frac{1}{N}\sum_{n=1}^{N}
\frac{1}{\lambda t^{*}}\sqrt{\frac{n}{2\pi}},\;\,\,N\to\infty \label{000}
\end{equation}
shows that the mass density at $X=M_{tot}/2\rho$ (\ref{000}) tends to infinity as $\sqrt{N}$. 

Equation (\ref{result}) is the main result of this Section. It tells us
that even in a nonstatic case where the aggregating particles have initially
a Gaussian velocity distribution the macroscopic mass is formed at the
gravitational time scale $t=t^{*}$. From this point of view the situation 
does not differ from the static problem discussed in Section 1.
However, the relation (\ref{tauzero}) clearly shows that the probability of
complete aggregation at $t=t^{*}$ is still zero. This  puts forward the
question of the distribution of those masses which have not joint 
the central macroscopic body at $t=t^{*}$.  The statistical
distribution of this leftover dust, representing a nonextensive amount
composed of microscopic fragments, is discussed in the next Section.
\\[0.7cm]

{\bf 3. STATISTICAL DISTRIBUTION OF MASSES}\\[0.5cm]

We first write down the probability distribution $$\mu_{k}^{N}(X_{1}V_{1}n_{1},\ldots,
X_{k}V_{k}n_{k};t)$$ for finding, at time $t$, an ordered configuration of $k$ aggregates 
at positions $X_{i},\; (X_{1}<X_{2}<\ldots <X_{k}$) with velocities $V_{i}$ and masses $M_{\n}=m\n$.

The aggregates originate from consecutive initial clusters made of $n_{1},\ldots, n_{k}$ initial masses
$m$ and are found at the center of mass of these clusters
\be
X_{\n}(t)=\left(n_{1}+n_{2}+\cdots +n_{i-1} +\frac{\n +1}{2}\right)a+\frac{V_{\n}}{\n}t+\gamma
m\Delta_{\n}\frac{t^{2}}{2}
\label{1}
\ee
with velocities
\be
V_{\n}(t)=\frac{V_{\n}}{\n}+\gamma m\Delta_{\n}t
\label{2}
\ee
The relations (\ref{1}) and  (\ref{2}) generalize  (\ref{cmnp}) and  (\ref{cmnv}) to the
splitting of the $N$ equidistant initial particles into $k$ consecutive clusters (here,
$i=1,\ldots,k$ does not index lattice sites, but the clusters themselves). In (\ref{1}) and (\ref{2}),
$V_{\n}$ is the sum of the initial velocities of the particles belonging to the $i^{th}$ cluster and
\be
m\Delta_{\n}=m(n_{i+1}+\cdots +n_{k}-n_{1}-\cdots -n_{i-1}) 
\label{3}
\ee
is the difference of the masses to the right and to the left of this cluster responsible for the force
acting on it.

The distribution $\mu_{k}^{N}$ is obtained by averaging the set of kinematical constraints required
for the realization of the desired event
\be
\mu_{k}^{N}(X_{1}V_{1}n_{1},\ldots,X_{k}V_{k}n_{k};t)=\label{4}
\ee
$$
\left\langle
\prod_{i=1}^{k-1}\theta(X_{i+1}-X_{i})\prod_{i=1}^{k}[\delta(X_{i}-X_{\n}(t))
\delta(V_{i}-V_{\n}(t))]\prod_{i=1}^{k}\Theta_{\n}(t)\right\rangle
$$

We always have $\sum_{i=1}^{k}n_{i}=N$, and the quantity $\Theta_{\n}(t)$ represents the constraint
(to be elaborated below) needed to ensure that the $i^{th}$ aggregate forms before time $t$. Both
$X_{\n}(t),V_{\n}(t)$ and $\Theta_{\n}(t)$ are known expressions of the initial  velocities and
$\langle\ldots\rangle$ denotes as before the mean value with respect to the distribution
(\ref{phiN}), so $\mu_{k}^{N}$ can be calculated in principle. Notice that we did not include further
constraints saying that the particles belonging to the adjacent clusters $i$ and $i+1$ do not perturb
each other before $t$. This is not needed since the very fact that all initial particles in the
$(i+1)^{th}$ cluster are found right of those in the $i^{th}$ cluster at time $t$ already implies
that no particles of the two groups have collisioned before $t$ (otherwise a double crossing between
trajectories would have occured, which is not possible in our dynamics).

To simplify the discussion, we shall merely be interested in the mass distribution by
integrating out positions and velocities
$$
\mu_{k}^{N}(n_{1},n_{2},\ldots,n_{k};t)
$$
\begin{eqnarray}
&=&\int dX_{1}\ldots dX_{k}\int dV_{1}\ldots dV_{k}
\mu_{k}^{N}(X_{1}V_{1}n_{1},\ldots,X_{k}V_{k}n_{k};t)\nonumber\\
&=&\left\langle
\prod_{i=1}^{k-1}\theta(X_{n_{i+1}}(t)-X_{\n}(t))
\prod_{i=1}^{k}\Theta_{\n}(t)\right\rangle
\label{5}
\end{eqnarray}
The conditions for forming the $i^{th}$ aggregate before $t$ are the same as (\ref{merge}), i.e.
$X_{\n}^{r}(t)>X_{\n}^{\n-r}(t),\; r=1, 2, \ldots, \n-1$, where $X_{\n}^{r}(t)$ is the position of the
subcluster of the first $r$ particles in the $i^{th}$ cluster
$$
X_{\n}^{r}(t)=\left(n_{1}+\cdots +n_{i-1}+\frac{r+1}{2}\right)a+\frac{1}{r}\sum_{s=1}^{r}v_{s}t\
$$
\be
+\gamma m(\n-r+n_{i+1}+\cdots +n_{k}-n_{1}-\cdots -n_{i-1})\frac{t^{2}}{2}
\label{6}
\ee
Here the velocities of the initial particles in this cluster have been simply labelled
$v_{1},v_{2},\ldots, v_{\n}$ and $V_{\n}=\sum_{s=1}^{\n}v_{s}$. Hence, as in (\ref{int}), the
constraint is 
\begin{eqnarray}
\Theta_{\n}(t)&=&\prod_{i=1}^{\n-1}\theta\left[X_{\n}^{r}(t)-X_{\n}^{\n-r}(t)\right]\nonumber\\
&=&\prod_{i=1}^{\n-1}\theta\left(\sum_{s=1}^{r}v_{s}-\frac{r}{\n}V_{\n}+\frac{r(\n-r)}{2}m\,\tau\right)
\label{7}
\end{eqnarray}
We see from (\ref{1}) that the factor 
\be
\prod_{i=1}^{k-1}\theta(X_{n_{i+1}}(t)-X_{\n}(t))=
\prod_{i=1}^{k-1}\theta\left(\frac{V_{n_{i+1}}}{n_{i+1}}-\frac{V_{n_{i}}}{n_{i}}
-\frac{n_{i}+n_{i+1}}{2}\,m\tau\right)
\label{8}
\ee
in (\ref{5}) depends only on the  initial center of mass velocities $V_{n_{i}}/{n_{i}}$ of the
clusters. 
Since the initial velocity distribution factorizes, we can perform the integration independently
for each cluster, except for the variables $V_{n_{i}}/{n_{i}}$ that are coupled through
(\ref{8}). Introducing for each cluster the change of variables (\ref{change}) one obtains as in
(\ref{int})-(\ref{P}) that the $i^{th}$ cluster contributes to the total integration on velocities
in (\ref{5}) as
$$
\int dv_{1}\ldots dv_{\n}\phi_{\lambda}(v_{1})\cdots\phi_{\lambda}(v_{\n})
\prod_{i=1}^{\n-1}\theta\left(\sum_{s=1}^{r}v_{s}-\frac{r}{\n}V_{\n}+\frac{r(\n-r)}
{2}m\,\tau\right)\ldots
$$
\be
=P_{\n}(\tau)\frac{1}{\sqrt{2\pi \n}\lambda}\int
dV_{\n}\exp\left(-\frac{V_{\n}}{2\n\lambda^{2}}\right)\ldots 
\label{9}
\ee
Taking (\ref{8}) and (\ref{9}) into account in (\ref{5}) and setting 
$U_{i}=V_{\n}/{\lambda\sqrt{\n}}$ leads to the final result 
$$
\mu_{k}^{N}(n_{1},n_{2},\ldots,n_{k};t)=\left(\prod_{i=1}^{k}P_{\n}(\tau)\right)
$$
\be
\times\int dU_{1}\ldots dU_{k}
\phi(U_{1})\ldots\phi(U_{k})\prod_{i=1}^{k-1}
\theta\left(\frac{U_{i+1}}{\sqrt{n_{i+1}}}-\frac{U_{i}}{\sqrt{\n}}-\frac{\n+n_{i+1}}
{2\lambda}\,m\tau\right)
\label{10}
\ee
The first factor is the probability of formation of independent aggregates, whereas the second
factor represents the correlations introduced between them by the gravitational forces.

We now draw some important conclusions from the formula (\ref{10}). We say that the $i^{th}$
aggregate is macroscopic if the number of its constituents $\n=\eta N, \;0<\eta\leq 1,$ is a non
vanishing fraction $\eta$ of the total number of initial particles as $N\to\infty$; its mass is then
$M_{i}=m\n=\eta M_{tot}$.

\vspace{3 mm}
(i) {\it Macroscopic mass}
\vspace{3 mm}

According to the form of the mass density (\ref{result}) at $t=t^{*}$, it is clear that there can be
only one macroscopic aggregate for $t>t^{*}$. Let us recover this result using (\ref{10}).
The arguments of the $\theta$-functions in (\ref{10}) are denoted by
\be
w_{i,i+1}=\frac{U_{i+1}}{\sqrt{n_{i+1}}}-\frac{U_{i}}{\sqrt{\n}}-\frac{\n+n_{i+1}}
{2\lambda}\,m\tau
\label{10a}
\ee
The integral in (\ref{10}) is carried out on the domain ${}\cal D$ of the variables
$U_{i}$ defined by $w_{i,i+1}\geq 0$. Suppose that the masses of the two clusters $j$ and $j+\ell$ 
become macroscopic, i.e. $n_{j}\to\infty$, $n_{j+\ell}\to\infty$ with $mn_{j}=M_{j}>0$,
$mn_{j+\ell}=M_{j+\ell}>0$ and $\n $ remains finite for $i\neq j, j+\ell$. Then as $N\to\infty$, 
\be
w_{j,j+1}=\frac{U_{j+1}}{\sqrt{n_{j+1}}}-\frac{M_{j}}{2\lambda}\,\tau
\label{11}
\ee
\be
w_{j+\ell-1,j+\ell}=-\frac{U_{j+\ell-1}}{\sqrt{n_{j+\ell-1}}}-\frac{M_{j+\ell}}{2\lambda}\,\tau
\label{12}
\ee
The relation (\ref{11}) together with the conditions  $w_{i,i+1}\geq 0$ for $\tau>0$ imply 
$U_{j+1}>0, \ldots , U_{j+\ell-1}>0$, but (\ref{12}) implies also $U_{j+\ell-1}<0$; thus the
integration domain ${\cal D}$ shrinks to zero as $N\to\infty$ and the corresponding probability
vanishes. The argument is the same when more than two masses become macroscopic.  

We calculate now the probability to have one macroscopic mass, say $n_{j}=N-\sum_{i\neq
j}^{k}{n_{i}}$, holding the other aggregates $n_{i}, \,i\neq j$ finite, defined by 
$$ 
\mu_{k}(n_{1},\ldots, n_{j-1},M_{tot},n_{j+1},\dots,n_{k};t)=
$$
\be
=\lim_{N\to\infty} \mu_{k}^{N}(n_{1},\ldots, n_{j-1},N-\sum_{i\neq
j}^{k}{n_{i}},n_{j+1},\dots,n_{k};t) \label{13}
\ee
Notice that in the
macroscopic limit, the mass $M_{n_{j}}=M_{tot}-m\sum_{i\neq j}^{k}n_{i}$ becomes
infinitesimally close to the total mass: $M_{n_{j}}\to M_{tot}$.  
We make two observations on the probability $P_{n}(\tau), \tau>0$. If $n$
is fixed
\be
\lim_{N\to\infty}P_{n}(\tau)=P_{n}(0)=\frac{1}{n}
\label{14}
\ee
since this amounts to let $m=M_{tot}/N\to 0$ in (\ref{P}). 
If $n=N-q$ with $q$ a fixed integer
\be
\lim_{N\to\infty}P_{N-q}(\tau)=P(\tau)
\label{15}
\ee
where
\begin{eqnarray}
P(\tau)&=&\exp(-A(\tau)),\nonumber\\
{\rm with}\;\;\; A(\tau)&=&2\sum_{n=1}^{\infty}\frac{1}
{\sqrt{n}}\int_{\frac{M_{tot}}{2\lambda}\,\tau}^{\infty}
\phi(\sqrt{n}y)dy \label{16} \end{eqnarray}
is the probability of merging of the total number of particles into the single mass $M_{tot}$.
This is precisely the function determined in the proposition found in section 5
of ref. [4], because again $m(N-q)=M_{tot}-mq\to M_{tot}$ as $N\to\infty$
\footnote{In this proposition the quantity $M_{tot}/{2\lambda}$ was set equal to one.}. 
Finally, as $N\to\infty$
the arguments $w_{i,i+1}$ (\ref{10a}) tend to
\begin{eqnarray}
w_{i,i+1}&=&\frac{U_{i+1}}{\sqrt{n_{i+1}}}-\frac{U_{i}}{\sqrt{n_{i}}}, \;\,\,\,i\neq j,\;\;i+1\neq j
\nonumber\\
w_{j,j+1}&=&\frac{U_{j+1}}{\sqrt{n_{j+1}}}-\frac{M_{tot}}{2\lambda}\;\tau
\nonumber\\
w_{j,j-1}&=&\frac{U_{j-1}}{\sqrt{n_{j-1}}}-\frac{M_{tot}}{2\lambda}\;\tau
\label{17}
\end{eqnarray}
When (\ref{14}), (\ref{15}) and (\ref{17}) are taken into account in (\ref{10}) (changing also
$U_{i}$ into $-U_{i}$), one finds that the limit (\ref{13}) is
$$ 
\mu_{k}(n_{1},\ldots, n_{j-1},M_{tot},n_{j+1},\dots,n_{k};t)=
$$
\be
=Q_{j-1}(n_{1},\ldots,n_{j-1};\tau)P(\tau)Q_{k-j}(n_{k},\ldots,n_{j+1};\tau)
\label{18}
\ee
with
$$
Q_{j-1}(n_{1},\ldots,n_{j};\tau)=
\frac{1}{n_{1}\cdots n_{j}}\int dU_{1}\ldots dU_{j}\phi(U_{1})\cdots \phi(U_{j})
$$
\be
\times\theta\left(\frac{U_{1}}{\sqrt{n_{1}}}-\frac{U_{2}}{\sqrt{n_{2}}}\right)\cdots 
\theta\left(\frac{U_{j-1}}{\sqrt{n_{j-1}}}-\frac{U_{j}}{\sqrt{n_{j}}}\right)
\theta\left(\frac{U_{j}}{\sqrt{n_{j}}}-\frac{M_{tot}}{2\lambda}\;\tau\right)
\label{19}
\ee
The interpretation of (\ref{19}) is clear: the factor $Q_{j-1}$ $(Q_{k-j})$ is the probability to 
find, left (right) of the macroscopic mass,
$j$ ($k-j$) aggregates made of a finite number of initial masses, that we call now fragments. We
shall show in paragraph (ii) below that the probabilities (\ref{18}) sum up to one. Therefore the
only configurations that can occur after $t^{*}$ consist of a single  macroscopic mass together with a
dust of such fragments. One should notice that non macroscopic pieces of matter of the order
$mN^{\nu},\;0<\nu<1$, do not appear after $t^{*}$.

\vspace{3 mm}
(ii) {\it Statistics of fragments}
\vspace{3 mm}

The probability to have a configuration of exactly k bodies after $t^{*}$ (i.e. one macroscopic
mass and $k-1$ fragments)
is
\be
\mu_{k}(t)=\sum_{j=1}^{k}\;\sum_{n_{1},\ldots,n_{j-1},n_{j+1},\ldots,n_{k}=1}^{\infty}
\mu_{k}(n_{1},\ldots, n_{j-1},M_{tot},n_{j+1},\dots,n_{k};t)
\label{20}
\ee
After the change of variables
$y_{i}=\left(\frac{U_{i}}{\sqrt{n_{i}}}-\frac{M_{tot}}{2\lambda}\;\tau\right)$, one finds from
(\ref{19})
\be
\sum_{n_{1},\ldots,n_{j}}^{\infty}Q_{j-1}(n_{1},\ldots,n_{j};\tau)\label{21}
\ee
\begin{eqnarray}
&=&\int_{y_{1\geq}\cdots y_{j-1}\geq y_{j}\geq
0}^{\infty}dy_{1}\ldots dy_{j}\:\prod_{i=1}^{j}\:\sum_{n=1}^{\infty}
\frac{1}{\sqrt{n}}\phi\left(\sqrt{n}\left[y_{i}+\frac{M_{tot}}{2\lambda}\,\tau\right]\right) 
\nonumber\\&=&\frac{1}{j\;!}\left(\frac{A(\tau)}{2}\right)^{j}\nonumber 
\end{eqnarray}
where $A(\tau)$ is the function defined in (\ref{16}). 
Hence from (\ref{18}), (\ref{20}), (\ref{21})
and (\ref{16}) one obtains the result 
\begin{eqnarray}
\mu_{k}(t)&=&P(\tau)\left(\frac{A(\tau)}{2}\right)^{k-1}\sum_{j=1}^{k}\frac{1}{(j-1)\:!(k-j)\:!}=
\nonumber\\
&=&\frac{\left(A(\tau)\right)^{k-1}}{(k-1)\;!}\exp(-A(\tau)),\;\;\;t\geq t^{*}
\label{22}
\end{eqnarray}
As claimed above, the $\mu_{k}(t)$ satisfy the normalization relation
$$\sum_{k=1}^{\infty}\mu_{k}(t)=1$$ The distribution of the number $k-1$ of fragments is Poissonian
for all times $t> t^{*}$. We find therefore that $A(\tau)$ appearing in (\ref{16}) has the
interpretation of the mean number of fragments. $A(\tau)$ tends to zero as a
Gaussian when $\tau\to\infty$ and diverges as $-2\ln\left(\frac{M_{tot}}{2\lambda}\,\tau\right)$
when $\tau\to 0$ (see eq. (52) in [4]).  

As a particular exemple, we write down from (\ref{18}) and (\ref{19}) the probability of survival 
of a fragment of size $n$
$$
\mu_{2}(M_{tot}, n;t)=\frac{1}{2}P(\tau)\left(1-{\rm
erf}\left(\frac{\sqrt{n}M_{tot}}{\sqrt{8}}\;\tau\right)\right)\sim
$$
\be
\sim\left\{\begin{array}{ll}\frac{1}{2}\left(\frac{M_{tot}}{2\lambda}\,\tau\right)^{2}& 
\mbox{if $\tau\to 0$}\\\frac{\sqrt{8}}{\sqrt{\pi n}M_{tot}\tau}
\exp\left(-\frac{n(M_{tot}\tau)^{2}}{8}\right)&\mbox{if $\tau\to \infty$}\end{array}\right.
\label{23}
\ee
where erf is the error function.

The general position and velocity dependent distributions (\ref{4}) for \\$t>t^{*}$ can also be
written down more explicitely in the macroscopic limit. In particular, one has
$\mu_{1}(X_{1}V_{1}M_{tot};t)=\delta(X_{1}-\frac{M_{tot}}{2\rho})\delta(V_{1})P(\tau)$. The
macroscopic mass is found at rest at the position $M_{tot}/{2\rho}$ without fluctuations, as
expected. More generally, there are Gaussian small probabilities to find fragments far away from this
point.

So far we have given a full description of the state after $t^{*}$: here the structure is simple
since the weight of typical mass configurations is given by the set of distributions (\ref{9})
with $k$ finite. When $t<t^{*}$ the situation is more complex. Indeed, we have
\be
\lim_{N\to\infty}\mu_{k}^{N}(n_{1},n_{2},\ldots,n_{k};t)=0,\;\,k=1,2,\ldots\;\,\,,t<t^{*}
\label{24}
\ee
since necessarily at least one of the $\n$ tends to infinity and
we know then from (\ref{tauzero}) that for 
$\tau\leq 0$, $P_{\n}(\tau)\leq 1/n \to 0$ (all the other factors in (\ref{10}) are bounded by
$1$). Hence there remain always infinitely many aggregates as $N\to\infty$, and the
weight of typical configurations will be given by the distributions
$\mu_{k}^{N}(n_{1}, \ldots,n_{k};t), \;k\to\infty$, involving infinitely many bodies. Computer
simulations indicate that after a short transient time, typical configurations consist of
approximately  $\sqrt{N}$ aggregates, each of them having a mass of the order $m\sqrt{N}$. Thus we
may conjecture that the distributions in this range, i.e.
$\sum_{n_{1},\ldots,n_{k}=c_{1}\sqrt{N}}^{c_{2}\sqrt{N}} \mu_{k}^{N}(n_{1},\ldots,n_{k};t)$
with $k\sim\sqrt{N}$, should have a non vanishing limit as $N\to\infty$. 

As far as the density (\ref{rhot}) is concerned, we anticipate that it converges for $t<t^{*}$ to an
absolutely continuous function, namely the uniform density (\ref{static}) as in the static model.
This would be consistent with the numerical observation that after a short time during which most of
the initial kinetic energy is dissipated by inelastic collisions, the subsequent evolution is
dominated by the gravitational forces. The result of simulations is discussed in the next Section.
Analytic proofs of these conjectures would complete the study of the dynamical phase transition that
occurs at $t^{*}$ between a spatially extended and an aggregated phase of matter.\\[0.7cm] 

{\bf 4. NUMERICAL SIMULATIONS}\\[0.5cm]

In this section we present the results of numerical simulations performed to determine   
the rate of formation of macroscopic masses for a system with a 
{\it finite} number $N$ of particles evolving according to the model described 
above. In particular, we analyze the scaling to the continuum limit of the  probability
$P^{\eta}_N(t)$  for the formation of a macroscopic mass $\eta N m,\;\; 0< \eta \leq 1$, before time
$t$.  We also study the time evolution of the kinetic energy. 

Numerical simulations on this model are particularly simple, compared with 
their counterparts in higher dimensions, because the equations of motion 
can be analytically integrated between succesive collisions. 
The simulation then reduces to keeping track of the particle masses, 
coordinates and velocities created in succesive collisions.

In going to the continuum limit while keeping the density constant, different 
scalings of the initial conditions are possible. 
Here, for reasons of numerical accuracy, we increase the number of 
particles $N$, while keeping the distance between them constant: $a=1$, and also putting $m=1$,
so that $\rho=M/L=nm/na=1$.
Further, we choose initial velocities with a Gaussian distribution of variance 
$\lambda=N/2$. This procedure is equivalent to the continuum limit considered in the preceding
sections (see (\ref{continuum}) and also [4]). 

In Fig. 1 we show the mass formation probability $P^{\eta}_N(t)$ for 
$N=1000$ as a function of time, averaged over 10000 initial configurations, with 
$\eta=$ 0.1, 0.2,..., 0.9, 0.99, 1.0 from left to right.

Note that $P^{\eta}_N(t)$ clusters around the Heaviside function for 
$\eta < 1$, while the probability of total mass aggregation $\eta=1$ 
follows the separate limiting curve given by (\ref{16}).

In order to study the scaling of $P^{\eta}_N(t)$ curves as a function of the particle 
number $N$, we obtained the results for the initial values  $N=10\cdot 2^r, 
r=0,1,\ldots,9$, averaged over $1000$ initial configurations.
Fixing $\eta=0.5$, we display in Fig. 2 the probability curves for 
increasing numbers of initial particles. They tend 
to the Heaviside function as $N\rightarrow \infty$.

Having in view a quantitative study of this scaling we consider the time deviation
$|t^{\eta}_{\beta}(N)-t^*|$  for increasing $N$, where $t^{\eta}_{\beta}(N)$ is the time  
at which the probability $P^{\eta}_N $ acquires the value $\beta_{i}$, i.e. $P^{\eta}_N
(t^{\eta}_{\beta}(N))=\beta,\;\;\;
\beta=0.25,0.5,0.75$, $\eta=0.5$. 

Fig. 3 reveals the power law 
$$ |t^{\eta}_{\beta}(N)-t^*|\simeq 
\frac{1}{\sqrt{N}}$$
The same behavior holds also for other values of $\eta$.

In order to get a deeper understanding of the dynamics of aggregation we analyze 
the evolution of the kinetic energy $E_{kin}^{N}(t)$. If we started from a static initial
configuration, we would simply find a parabolic law
$$E^{N,stat}_{kin}(t)=\frac{\gamma^2}{6}M_{tot}^3t^{2}$$
In Fig. 4 the ratio $E_{kin}^{N}(t)/E^{N,stat}_{kin}(t^{*})$ has been plotted. 
This ratio approaches the normalized parabola as $N\to\infty$. One can look at the local minimum of
the curves as corresponding to the time scale of almost total dissipation of the kinetic energy due
to initially numerous inelastic collisions. For $N\to\infty$, the location of the minimum 
approaches the initial moment $t=0$ according to the power law $N^{-1/4}$. Hence in the continuum
limit the system gets instantaneously cooled down and the subsequent evolution is dominated by
gravity.

We also observe that at the time when the kinetic energy attains its minimum:
\begin{description}
\item[(i)] the average size of the formed masses scales as $\sim m\sqrt{N}$
\item[(ii)] the velocity distribution remains 
Gaussian with an effective standard deviation $\lambda_{eff}\sim \lambda N^{-1/4}$.
\end{description}
It strongly suggests that before $t^{*}$ the density of mass in the continuum limit should also
coincide with that of the static model (\ref{static}).

Finally, in order to study the sensitivity of our results to variations of the 
initial distributions, we have led the same simulations for 
interparticle distances following a Poissonian distribution and a Gaussian 
initial velocity distribution. We confirmed that all the results presented 
above remained identical, indicating a certain generality of the initial
kinetic energy dissipation process.\\[0.7cm]

{\bf ACKNOWLEDGEMENTS}\\[0.5cm]

J. Piasecki acknowledges the hospitality at Institut de Physique Th\'eorique de 
l'Ecole Polytechnique F\'ed\'erale de Lausanne, and partial financial support by
KBN (Committee for Scientific Research, Poland), grant 2 P03B 03512.
This work was supported by the Swiss National Fond Grants No. 20-39528.93,
and the University of Fribourg.\\[0.7cm]

{\bf REFERENCES}\\[0.5cm]
1. Ya. B. Zeldovich, {\it Astron. Astrophys.} {\bf 5}: 84 (1970).\\
2. M. Vergassola, B. Dubrulle, U. Frisch and A. Noullez, {\it Astron. Astrophys.}
   {\bf 289}: 325 (1994).\\
3. E. Weinan, Yu.G. Rynov, Ya.G. Sinai, {\it Commun.Math.Phys.}
   {\bf 177}:349 (1996).\\
4. Ph.A. Martin, J. Piasecki, {\it J.Stat.Phys.} {\bf 84}:837 (1996).\\
5. Ph.A.  Martin, J. Piasecki, {\it J.Stat.Phys.} {\bf 76}:447 (1994).\\
6. W. Feller, {\it An Introduction to Probability Theory and Its
   Applications} (Wiley, New York, 1968).

\newpage

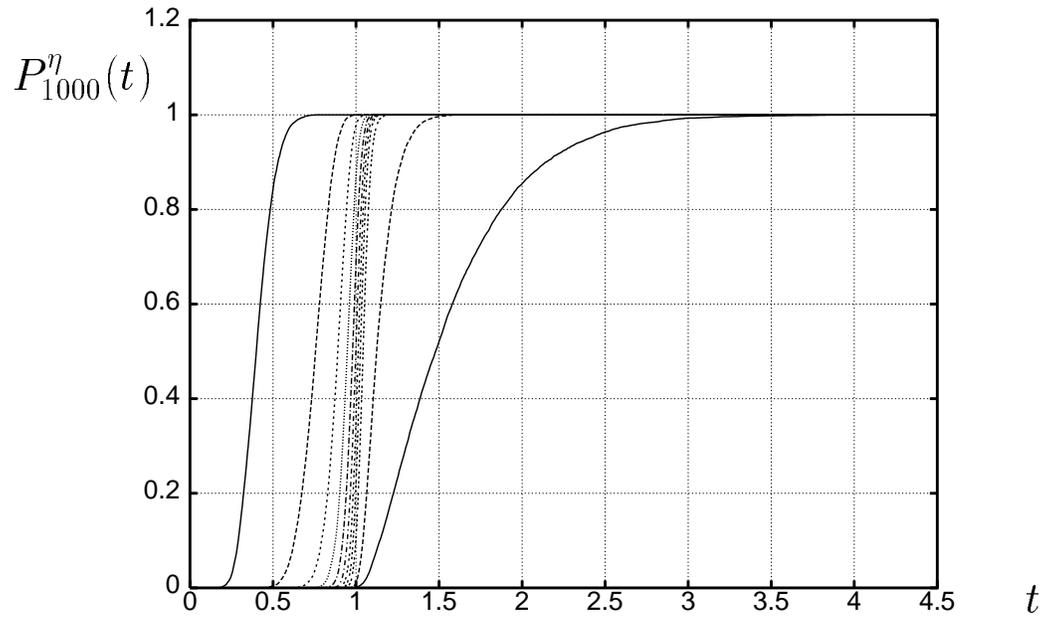
Fig. 1: Mass formation probability $P^{\eta}_N(t)$, N=1000, 
$\eta=$ 0.1, 0.2,..., 0.9, 0.99, 1.0 from left to right.
\vspace{0.5cm}

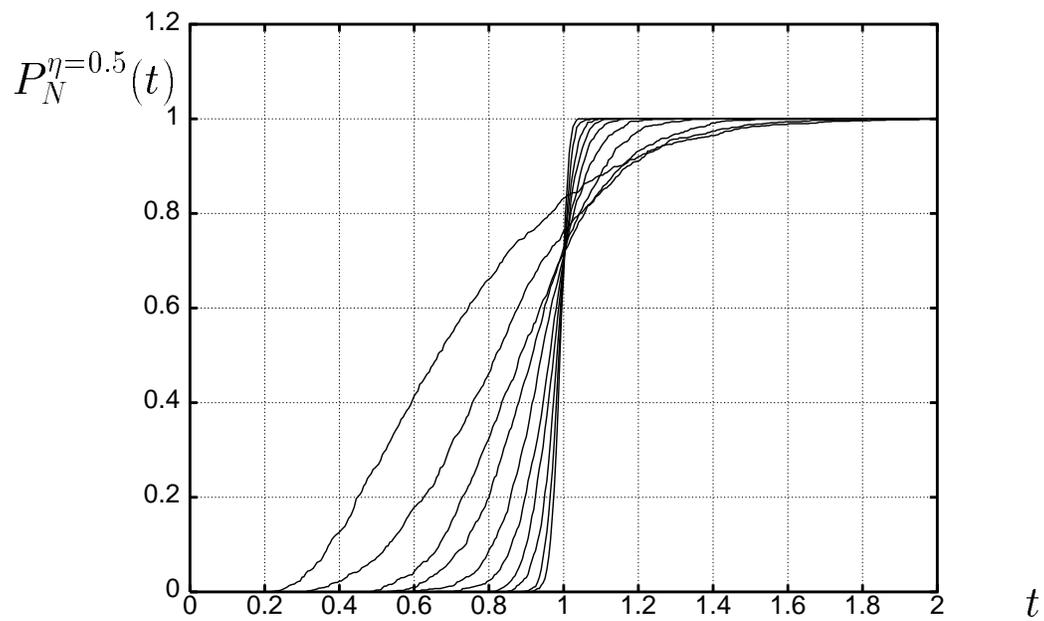
Fig. 2: Mass formation probability $P^{0.5}_N(t)$,  $N=10\cdot 2^r,r=0,1,\ldots,9$, 
from left to right.
\vspace{0.5cm}

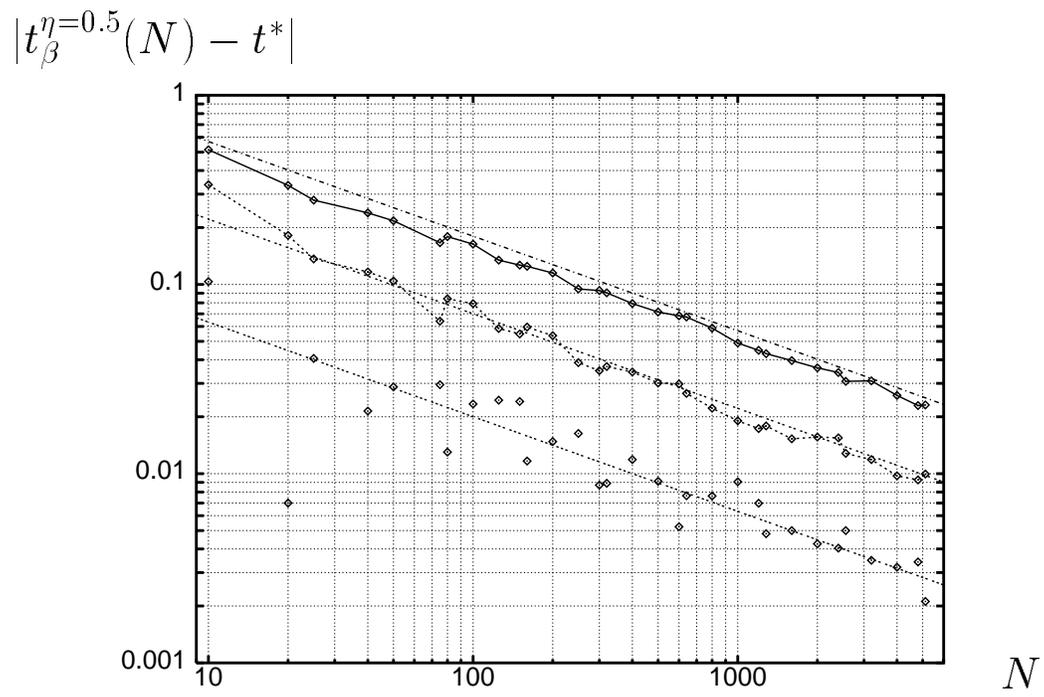
Fig. 3: Log-log plot of $|t^{\eta}_{\beta}(N)-t^*|$, for 
$\beta=0.25,0.5,0.75$ from up to down.
\vspace{0.5cm}

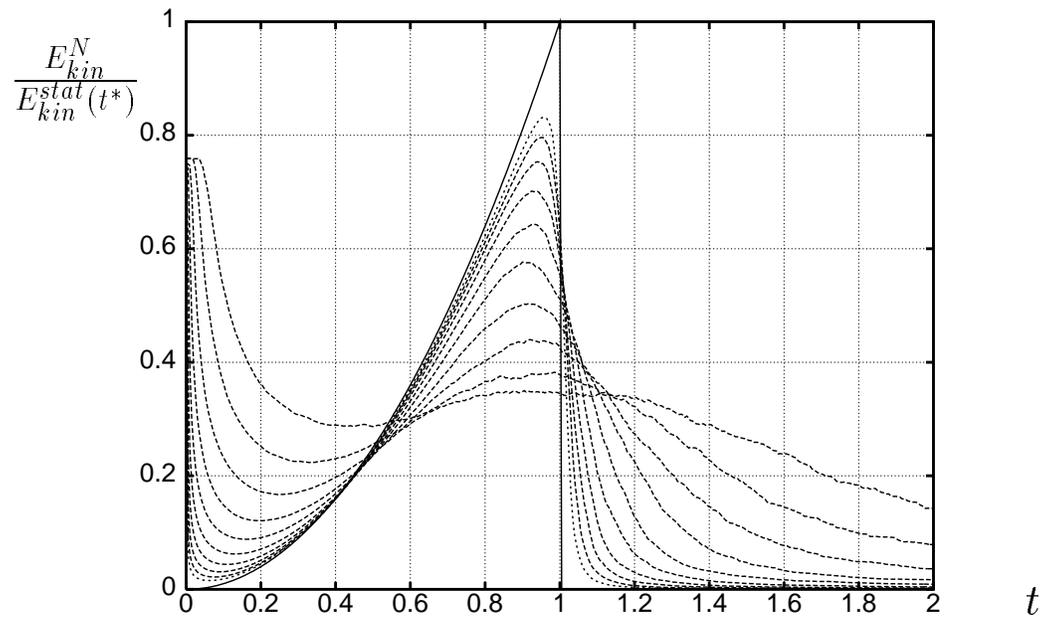
Fig. 4: Normalized kinetic energy as a function of time for\\ 
$N=10\cdot 2^r,r=0,1,\ldots,9$. The parabolic limit curve represents the normalized 
kinetic energy of the continuous static initial configuration. 

\end{document}